%% file: main.tex
\documentclass[aps,prd,superscriptaddress,onecolumn,showpacs,nofootinbib,preprintnumbers]{revtex4}

\usepackage{amsmath,amssymb,hyperref,subfigure,xspace}
\usepackage{mciteplus,color,mathtools,graphicx}
\usepackage[english]{babel}
\usepackage{todonotes}
\usepackage[percent]{overpic}
\usepackage{amsmath}
\include{review_tool}

\include{notations}

\usepackage{tikz}
\input{mytikzset.tex}
\input{tikzdiagrams.tex}

\begin{document}
\preprint{JLAB-THY-19-2924}

\title{Three-body scattering: Ladders and Resonances}

\allowdisplaybreaks

\newcommand{\ceem}{Center for  Exploration  of  Energy  and  Matter,
Indiana  University,
Bloomington,  IN  47403,  USA}
\newcommand{\indiana}{Physics  Department,
Indiana  University,
Bloomington,  IN  47405,  USA}
\newcommand{\jlab}{Theory Center,
Thomas  Jefferson  National  Accelerator  Facility,
Newport  News,  VA  23606,  USA}
\newcommand{\hiskp}{Universit\"at Bonn,
Helmholtz-Institut f\"ur Strahlen- und Kernphysik, 53115 Bonn, Germany}
\newcommand{\ect}{European Centre for Theoretical Studies in Nuclear Physics and related Areas (ECT$^*$) and Fondazione Bruno Kessler, Villazzano (Trento), I-38123, Italy}
\newcommand{\genova}{INFN Sezione di Genova, Genova, I-16146, Italy}
\newcommand{\cern}{CERN, 1211 Geneva 23, Switzerland}
\newcommand{\ucm}{Departamento de F\'isica Te\'orica, Universidad Complutense de Madrid, 28040 Madrid, Spain}

\author{M.~Mikhasenko}
\email{mikhail.mikhasenko@cern.ch}
\affiliation{\hiskp}
\affiliation{\cern}

\author{Y.~Wunderlich}
\email{wunderlich@hiskp.uni-bonn.de}
\affiliation{\hiskp}

\author{A.~Jackura}\affiliation{\ceem}\affiliation{\indiana}
\author{V.~Mathieu}\affiliation{\jlab}
\affiliation{\ucm}

\author{A.~Pilloni}
\affiliation{\ect}
\affiliation{\genova}

\author{B.~Ketzer}
\affiliation{\hiskp}

\author{A.~P.~Szczepaniak}
\affiliation{\ceem}\affiliation{\indiana}\affiliation{\jlab}

\begin{abstract}
We discuss unitarity constraints on the dynamics of a system of three interacting particles.
We show how the short-range interaction that describes three-body resonances can be separated from
the long-range exchange processes, in particular the one-pion-exchange process.
It is demonstrated that unitarity demands
a specific functional form of the amplitude with a clear interpretation:
the bare three-particle resonances are dressed by the initial- and final-state interaction, in a way that is consistent with the considered long-range forces.
We postulate that
the resonance kernel admits a factorization in the energy variables of the initial- and the final-state particles.
The factorization assumption leads to an algebraic form for the unitarity equations, which is reminiscent of the well-known two-body-unitarity condition and approaches it in the limit of the narrow-resonance approximation.
\end{abstract}

\maketitle

%

\section{Introduction} \label{sec:Intro}
Unitarity of the $S$-matrix
is one of the most important constraints on reaction amplitudes.
On one hand, it ensures probability conservation in scattering reactions,
on the other, when combined with analyticity~\cite{Eden:1966dnq}, it enables to correlate structures in measured cross sections with properties of the underlying resonances.
The vast majority of hadron resonances can be classified by their valence quark (antiquark) content.
A much richer spectrum, however, is expected from Quantum Chromodynamics (QCD) and, consequently, QCD dynamics cannot be fully understood without exploring hadron structure beyond the quark model.
Recent experiments have identified numerous signatures of such exotic states
as, for example, $XYZ$ states in the charmonium sector~\cite{Olsen:2017bmm,Esposito:2016noz,Guo:2017jvc},
and $\pi_1(1600)$, $\pi_2(1880)$, and $a_1(1420)$ in the light meson sector~\cite{Akhunzyanov:2018lqa,Adolph:2015pws}.
In parallel, great progress has been made in calculating hadron reactions from first principles, in lattice QCD~\cite{Dudek:2012xn,Dudek:2014qha,Briceno:2016mjc,Briceno:2017max}.
Since the majority of these new states is observed decaying to three particles, it is necessary to develop \threeToThree scattering amplitudes that satisfy the $S$-matrix principles and therefore can be used to extract resonances from the analyses of the experimental and lattice simulation data~\cite{Hansen:2016ync,Mai:2017vot,Jackura:2018xnx}.
Furthermore, a better understanding of three-hadron dynamics is necessary to take full advantage of the large data sets from COMPASS, LHCb, CLAS, GlueX, BESIII, and BelleII
where genuine three-body effects have already been observed
(\eg see the $a_1(1420)$ phenomenon~\cite{Krinner:2017vch,Adolph:2015pws,Ketzer:2015tqa,Aceti:2016yeb}).

The goal of this paper is to present an approach to construct the \threeToThree scattering amplitude within the isobar representation, which satisfies unitarity and analyticity of partial waves by separating the long-range interactions from the short-range QCD dynamics.
The \textit{isobar representation}, when dealing with three particles, is a form of the amplitude written as a sum of three partial-wave series, one for every pair of particles~\cite{Herndon:1973yn}.
An alternative, \textit{partial-wave representation} of the scattering amplitude corresponds to projecting the amplitude onto a single set of partial waves with a particular choice of the third particle. In the partial-wave representation, in order to reproduce the threshold singularities in all two-body subchannels, it would be necessary to
explicitly include an infinite number of partial waves.
The isobar representation has the advantage that such singularities can be accounted for even with a finite number of isobar terms. Moreover, the isobar representation naturally incorporates the permutation symmetry required when dealing with indistinguishable particles.

In our model, we explore the idea separating the scattering amplitude components according to the different interaction ranges.
The long-range interaction is dominated
by one-pion-exchange (OPE) processes and is inherent to the three-particle dynamics.
The short-range interaction is expected to govern resonance formation.
Unitarity in three-particle systems has been studied extensively in the past~\cite{Grisaru:1966xyz,Mandelstam:1962xyz,Harrington:1962pr,Fleming:1964zz,Frazer:1962pr,Holman:1965prb,Cook:1962zz,Ball:1962bwa,Hwa:1964xyz,Eden:1966dnq}.
For the derivation of the unitarity equations,
we follow the work of G.~Fleming~\cite{Fleming:1964zz}. However, the model which we deduce from said equations has a major difference compared to Ref.~\cite{Fleming:1964zz, Holman:1965prb}. In the latter, discontinuity relations for the  isobar-spectator partial waves were derived in all relevant variables in order to formulate the $N/D$ equations~\cite{Bjorken:1960prl}.
Schematically, the $D$ function is constructed from a driving term $N$ in such a way that the full amplitude satisfies unitarity.
In contrast, in our approach we identify an analytical solution of the unitary relations without solving the complicated $N/D$ equations.
Specifically, for the long-range part one solves a linear integral equation of the Blankenbecler and Sugar (BS) type~\cite{Blankenbecler:1965gx}
(for application to the three-body problem see Ref.~\cite{Amado:1975zz,Amado:1974za,Aaron:1973ca,Aaron:1969my,Mai:2017vot}),
while the short-range interactions are incorporated additively.
As we demonstrate, unitarity requires to append
an infinite series of exchange processes to both sides of the short-range kernel.
Such an infinite series of rescatterings is provided implicitly by the solutions of the well-known Khuri-Treiman (KT) equations~\cite{Khuri:1960zz,Pasquier:1968zz,Pasquier:1969dt,Bronzan:1963mby,Aitchison:1965zz,Aitchison:1966lpz,Kacser:1966jmp,Aitchison:1978pw,Aitchison:1979fj,Aitchison:1979ja}. The KT framework is based on the assumption of the analytic continuation of two-body unitarity equation from the scattering domain to the three-particle decay region.
The KT framework provides
a quantitative estimate for the size of rescattering corrections.
As it turns out,
these corrections are the only ingredients that enter the unitarity constraint to the properties of resonances.
In Ref.~\cite{Aitchison:1966lpz,Pasquier:1968zz} it was shown that KT equations contain a particular realization of the \threeToThree dynamics.
We adopt this model and matched with the three-body unitarity requirements.

The central idea of the paper is the factorization assumption. Guided by the factorization of residues at the resonance pole,
we assumed that the short-range kernel can be written as a product of functions
that depend on either energy variables of the incoming particles or the outgoing ones.
The full \threeToThree scattering amplitude does not factorize due the one-pion-exchange process.
However, as soon as the short-range part of amplitude is considered separately, the factorization ansatz can be implemented consistently.
Under this assumption, unitarity becomes an algebraic constraint
that is reminiscent of the two-body-unitary condition.

The paper is organized as follows.
In Sec.~\ref{sec:Model} we present a general model consistent with three-body unitary that treats separately the long- and short-range interactions.
We show that the scattering amplitude can be written as a sum of a function $\Ladder$ representing the ladder of exchanges
and another one, $\Rcal$, which contains the resonance physics dressed by initial- and final-state interactions.
In Sec.~\ref{sec:factorization}, the factorization assumption is discussed. Following the simplification of the constraints,
we derive two practical limits when the subchannel interaction is described by a narrow resonance.
Sec.~\ref{sec:SpecificModel} is dedicated to the relation of the three-body-unitarity and Khuri-Treiman model.
We discuss a three-particle-production amplitude and a specific realization of the long-range kernel.
Conclusions and outlook are given in Sec.~\ref{sec:ConclusionsAndOutlook}.

%

\section{A general unitary model}
\label{sec:Model}

In order to simplify the presentation, we consider the case of three identical, scalar particles with mass $\mpi$ for which the interaction is only significant in the $S$-wave.
In our simple setup we do not consider the isospin symmetry assuming that the generalization is straightforward and is required to be done when higher partial waves are included.
The scattering amplitude is defined as the expectation value of the transition operator $T$ sandwiched
between projected three-particle states as discussed in Appendix~\ref{app:derivation.of.unitarity}.
To derive the isobar representation,
the symmetrized three-particle state is decomposed into
three series of partial-wave-projected states.
The partial-wave projection of the three-particle state is done in two steps.
First, the state of a selected pair of particles is expanded in the helicity basis.
Second, the partial-wave expansion is performed in the overall center-of-mass frame
combining the particle-pair state (\textit{isobar}), and the remaining third particle (\textit{bachelor}).
The interaction operator is split into the fully \textit{connected} $\Tc$ and the partially \textit{disconnected} $\Td$
following the connectedness principle of Ref.~\cite{Eden:1966dnq}.
An equivalent separation is obtained using the LSZ reduction~\cite{Itzykson:1980rh,Lehmann:1954rq}.
We write, schematically,
\begin{align} \label{eq:general.isobar.model}
  \tikzdiagThreeToThree &\,=\, \sum_{9}\left(3\,\tikztauUpUp + \tikzconUpUpFULL\right)
\end{align}
with $T = \sum_{Q=1}^3 \Td^Q + \Tc$, where
the sum for the identical particles acting on different subchannels gives
a factor of $3$ for the disconnected part (for details see Appendix~\ref{app:derivation.of.unitarity}, also discussions in Ref.~\cite{Eden:1966dnq,Fleming:1964zz}).
The sum over $9$ combinations count all possible ways of choosing the third particle in the initial and final state.
Each of the nine connected terms in Eq.~\eqref{eq:general.isobar.model} is considered as a partial-wave series.

The $S$-wave projection of the two-particle scattering amplitude is denoted by $t(\sigma)$, where $\sigma$ is the invariant mass squared of these two particles. For the $S$-wave amplitude, each of the nine connected amplitudes in Eq.~\eqref{eq:general.isobar.model} is given by the same  scalar function of the three-particle invariant mass squared, $s$ and the two isobar invariant masses squared in the initial and final state, respectively. Schematically,
\begin{align}
  \tikztauUpUp\, &= t(\sigma),\\ \label{eq:amputated}
  \tikzconUpUpFULL \equiv \tikztauUpUp\tikzconUpUp\tikztauUpUp &= t(\sigma')\,\Tcal(\sigma',s,\sigma)\,t(\sigma),
\end{align}
By separating the $t(\sigma^{(\prime)})$ functions on both sides, we define the amputated isobar amplitude which is indicated by the large empty circle in the drawing. As shown in Appendix~\ref{app:derivation.of.unitarity}, the unitarity equation is simpler for the amputated isobar amplitude  than for the fully connected amplitude.
The left-hand singularities of $t(\sigma)$ are here ignored, but they could be taken care of by replacing $t(\sigma)$ in Eq.~\eqref{eq:amputated} with the \Omnes function~\cite{Omnes:1958hv}.

The unitarity condition, that is $T-T^\dagger = iT^\dagger T$ for the abstract interaction operator, splits into two separate equations: one for the connected amplitude, and another one for the disconnected amplitude.
The condition for the disconnected part reproduces the unitarity relation for the $2\to 2$ scattering amplitude $t(\sigma)$:
\begin{align}\label{eq:unitarity.t}
  \tikztauUpUp - \tikztauUpUp\,^\dagger & &
  t(\sigma) - t^\dagger(\sigma) &= i\,t^\dagger (\sigma) \rho(\sigma) t(\sigma)\,\theta(\sigma-4\mpisq).
  &\tikztauUpUp&\tikzCUpUp\tikztauUpUp
\end{align}
Here, the two-body phase-space factor is given by $\rho(\sigma) = \lambda^{1/2}(\sigma,\mpisq,\mpisq)/(8\pi\sigma)$. The K\"{a}ll\'{e}n triangle function is defined as $\lambda(x,y,z) = x^2+y^2+z^2-2xy-2yz-2zx$ \cite{BycklingKajantie:1971}.
The amplitude $t^\dagger(\sigma) \equiv t^*(\sigma)$ is defined as the expectation value of the operator $T^\dagger$ between the $S$-wave projected two-particle states.
Similarly, $\Tcal^\dagger(\sigma',s,\sigma) =\Tcal^*(\sigma,s,\sigma')$. Moreover, for strong interactions time-reversal symmetry implies $\Tcal(\sigma',s,\sigma) =\Tcal(\sigma,s,\sigma')$.
The three-body unitarity equation for the reduced $S$-wave amplitude $\Tcal (\sigma',s,\sigma)$  then reads (\cf Appendix \ref{app:derivation.of.unitarity}):
\parbox{\textwidth}{
\begin{subequations}\label{eq:three.body.unitarity}
  \begin{align}\nonumber
  &\Tcal(\sigma',s,\sigma)-\Tcal^\dagger(\sigma',s,\sigma) =
  \\
  \label{eq:unitary.line4}
    &\qquad 2i\, \frac{1}{\lamHs(\sigma')} \frac{1}{8\pi}\int_{\sigma^-(\sigma',s)}^{\sigma^+(\sigma',s)} \diff\sigma_3'\,t(\sigma_{3}') \Tcal(\sigma_{3}',s,\sigma)
    && \tikzDnUp\tikztauUpUp\tikzconUpUp\\
  \label{eq:unitary.direct}
  &\qquad +\frac{i}{3}\,\int_{4\mpisq}^{(\sqrt{s}-\mpi)^2}  \frac{\diff \sigma''}{2\pi} \Tcal^\dagger(\sigma',s,\sigma'') \,t(\sigma'')t^\dagger(\sigma'') \,
    \rho(\sigma'')\rho_s(\sigma'')\,\Tcal(\sigma'',s,\sigma)
    &\tikzconUpUp\tikztauUpUp&\tikzCUpUp\tikztauUpUp\tikzconUpUp \\
  \label{eq:unitary.recoupling}
  &\qquad + \frac{2i}{3}\,\frac{1}{(8\pi)^2}\iint_{\phi(\sigma_2'',s,\sigma_3'')>0}  \frac{\diff \sigma_2''\diff \sigma_3''}{2\pi s}\,
          \Tcal^\dagger\,(\sigma',s,\sigma_{2}'') t^\dagger\,(\sigma''_{2})t(\sigma''_{3}) \Tcal(\sigma_{3}'',s,\sigma)
  &\tikzconUpDn\tikztauDnDn&\tikzDnUp\tikztauUpUp\tikzconUpUp \\
  \label{eq:unitary.line5}
  &\qquad + 2i\, \frac{1}{\lamHs(\sigma)} \frac{1}{8\pi}\int_{\sigma^-(\sigma,s)}^{\sigma^+(\sigma,s)} \diff\sigma_2\, \Tcal^\dagger(\sigma',s,\sigma_{2}) t^\dagger(\sigma_{2})
  &\tikzconUpDn\tikztauDnDn&\tikzDnUp \\
  \label{eq:unitarity.ope}
  &\qquad + 6i\,\frac{2\pi s}{\lamHs(\sigma')\lamHs(\sigma)} \,\theta^+(\phi(\sigma',s,\sigma)).
  && \tikzDnUp
  \end{align}
\end{subequations}
}
Here, $\rho_s(\sigma) = \lamHs(\sigma)/(8\pi s)$ with $\lambda_s(\sigma) \equiv \lambda(s,\sigma,\mpisq)$ describes the isobar-bachelor phase-space factor.
The term in Eq.~\eqref{eq:unitary.direct} gives a direct coupling of the particles combined into the isobar, while the term~\eqref{eq:unitary.recoupling} involves the \textit{recoupling},
\ie the pairs of particles coupled to an isobar are different for the amplitudes $\Tcal$ and $\Tcal^{\dagger}$.
A term in Eq.~\eqref{eq:unitarity.ope} represents a real-particle exchange between isobars in the initial and final state, which is kinematically allowed only in the \textit{decay region}, \ie the exchanged particle can only be on its mass shell
  when, for given $s$, $\sigma$ and $\sigma'$ are inside the kinematic limits of the Dalitz plot.
Therefore, the \textit{decay region} is defined by the condition $\phi(\sigma',s,\sigma) > 0$,
with $\phi = \sigma\sigma'(3\mpisq+s-\sigma-\sigma')-\mpisq(s-\mpisq)^2$ being the Kibble function~\cite{Kibble:1960zz},
together with $s>9\mpisq$ and $\sigma^{(\prime)}>4\mpisq$.
The function $\theta^+(\phi(\sigma',s,\sigma))$ combines four Heaviside functions that implement these restrictions (\cf Eq.~\eqref{eq:Heaviside.plus}).
For fixed $s$ and $\sigma'$, the integration limits in Eq.~\eqref{eq:three.body.unitarity} are determined by the Dalitz plot boundary, $\phi(\sigma',s,\sigma) =0$, which yields $\sigma^\pm(s,\sigma) = (s+3\mpisq-\sigma)/2 \pm \lamHs(\sigma)\lambda^{1/2}(\sigma)/(2\sigma)$ as solutions.
To simplify the further discussion, we introduce a shorthand notation,
in which the integrations over the $\sigma$-variables are implicit.
In order to achieve this, we unify the integration limits for all integrals in Eq.~\eqref{eq:three.body.unitarity} with use of Heaviside functions.
In our shorthand notation, Eq.~\eqref{eq:three.body.unitarity} reads
\begin{equation} \label{eq:three.body.unitarity.short}
  \Tcal - \Tcal^\dagger = \dB\tau\Tcal + \Tcal^\dagger(\tau-\tau^\dagger)\Tcal + \Tcal^\dagger\tau^\dagger\dB\tau\Tcal + \Tcal^\dagger\tau^\dagger\dB + \dB,
\end{equation}
where we defined $\tau(\sigma) = t(\sigma)\rho_s(\sigma)/3$ and
\begin{equation} \label{eq:dB.def}
  \dB \left( \sigma', s, \sigma \right)  =  2\pi i\, \frac{6 s}{\lamHs(\sigma')\lamHs(\sigma)} \,\theta^+(\phi(\sigma',s,\sigma)).
\end{equation}
A multiplication by $\tau$ in the shorthand notation implies an integral over a sub-energy variable
$\sigma$, shared by $\tau$ and two functions on both sides of it in the product:
\begin{equation}  \label{eq:TauMultRule}
X \tau Y \longleftrightarrow \int_{4m_\pi^2}^{( \sqrt{s} - \mpi )^{2}} \frac{\diff\sigma}{2\pi}\, X ( \ldots,s, \sigma ) \rho_s (\sigma) t (\sigma) Y ( \sigma, s,\ldots ).
\end{equation}
Note that we used Eq.~\eqref{eq:unitarity.t} to identify the term in Eq.~\eqref{eq:unitary.direct} with the second term on the right side of Eq.~\eqref{eq:three.body.unitarity.short}.
Interestingly, the lower integration limit in Eq.~\eqref{eq:TauMultRule} can be chosen smaller then $4\mpisq$.
The reason for this appearing arbitrariness is given by the fact that the integration limits are truncated to the physical region, ie $\sigma^{(')}>4\mpisq$, $\sqrt{s}>9\mpisq$ and $\phi(\sigma',s,\sigma)>0$,
by the Heaviside functions in every term of Eq.~\eqref{eq:three.body.unitarity.short}.

It is important to realize that Eq.~\eqref{eq:three.body.unitarity.short} does not specify a model for the scattering amplitude unambiguously.
It is a constraint that has to either be checked for whatever model is considered, or, built in explicitly during the construction of the model.
For our model construction, we follow the latter approach.
As a first step, we decompose $\Tcal (\sigma', s, \sigma)$ into a sum of a
\textit{long-range} piece, described by what we call the \textit{ladder} amplitude $\Ladder$, and the  \textit{short-range} term, which we refer to by $\Rcal$,
\begin{align}\label{eq:L+R}
  \tikzconUpUp && \Tcal(\sigma',s,\sigma) &= \Ladder(\sigma',s,\sigma)+\Rcal(\sigma',s,\sigma).&&\tikzLUpUp + \tikzRUpUp
\end{align}
There are no assumptions implied in the decomposition, however, it will lead to a clear intuitive picture of the interactions.
In the remainder of this section, we will give meaning to the functions $\Ladder$ and $\Rcal$ one after the other and demonstrate how a unitary model for the scattering amplitude emerges.

We use a general ansatz, the BS-type equation, 
\footnote{
  In Ref.~\cite{Mai:2017vot}, this equation is referred to as Bethe-Salpeter equation. We find it more appropriate
  to name it Blankenbecler-Sugar equation following Ref.~\cite{Blankenbecler:1965gx}, where a similar form of equation was proposed in the context of the $S$-matrix approach.
} which satisfies the three-body-unitarity constraint in Eq.~\eqref{eq:three.body.unitarity.short} as suggested in Ref.~\cite{Mai:2017vot}.
The ladder amplitude, $\Ladder$ is defined via iterating the long-range kernel $\Ker \left( \sigma', s,\sigma \right)$:
\begin{align} \label{eq:ladder.definition}
  \tikzLUpUp && \Ladder    & = \Ker + \Ladder\tau\Ker = \Ker + \Ker \tau \Ladder. && \tikzBUpDn+\tikzLUpUp\tikztauUpUp\tikzBUpDn
\end{align}
The kernel $\Ker$ satisfies
\begin{equation} \label{eq:BKernelDisc}
\Ker-\Ker^\dagger = \dB,
\end{equation}
where $\dB$ is the discontinuity of the real-particle exchange introduced in Eq.~\eqref{eq:dB.def}.
This condition does not specify $\Ker$ uniquely (see discussion on different choices in Ref.~\cite{Jackura:2018xnx}).
The discussion in the rest of this section is general, while we will choose a specific form of the kernel in Sec.~\ref{sec:SpecificModel}.

The unitarity of the amplitude $\Ladder$ given by Eq.~\eqref{eq:ladder.definition} is proven already in Ref.~\cite{Mai:2017vot} formally inverting the operators.
Using the short notation from Eq.~\eqref{eq:TauMultRule}, the linear proof becomes straightforward:
\begin{align} \nonumber
  \Ladder - \Ladder^\dagger &= \Ladder^\dagger(\tau-\tau^\dagger)\Ladder + (1+\Ladder^\dagger\tau^\dagger)\Ladder - \Ladder^\dagger(1+\tau\Ladder) \\ \nonumber
                        &= \Ladder^\dagger(\tau-\tau^\dagger)\Ladder + (1+\Ladder^\dagger\tau^\dagger)\Ker(1+\tau\Ladder) - (1+\Ladder^\dagger\tau^\dagger)\Ker^\dagger(1+\tau\Ladder)\\ \nonumber
                        &= \Ladder^\dagger(\tau-\tau^\dagger)\Ladder + (1+\Ladder^\dagger\tau^\dagger)\,\dB\,(1+\tau\Ladder)\\
                        &= \dB\tau\Ladder + \Ladder^\dagger(\tau-\tau^\dagger)\Ladder + \Ladder^\dagger\tau^\dagger\dB\tau\Ladder + \Ladder^\dagger\tau^\dagger\dB  + \dB. \label{eq:Ladder.unitarity}
\end{align}
were the five terms in Eq.~\eqref{eq:Ladder.unitarity} directly
correspond to the five terms in Eq.~\eqref{eq:three.body.unitarity.short}.
Inserting Eq.~\eqref{eq:L+R} into the unitarity equation~\eqref{eq:three.body.unitarity.short} and eliminating contributions from the ladder (\cf Eq.~\eqref{eq:Ladder.unitarity}), we obtain the unitarity constraint for $\Rcal$, which reads
\begin{subequations} \label{eq:unitarity.pw.R}
\begin{align}
  \label{eq:unitarity.pw.R.disc.s}
  \Rcal - \Rcal^\dagger &= \Rcal^\dagger(\tau-\tau^\dagger)\Rcal + \Rcal^\dagger \tau^\dagger \dB \tau \Rcal
&\tikzRUpUp\tikztauUpUp\tikzCUpUp\tikztauUpUp\tikzRUpUp\, + \, \tikzRUpDn\tikztauDnDn\tikzDnUp\tikztauUpUp\tikzRUpUp &\\
   \label{eq:unitarity.pw.R.disc.sigma.prime}
    &\qquad +\big(\dB\tau + \Ladder^\dagger(\tau-\tau^\dagger) + \Ladder^\dagger \tau^\dagger \dB \tau \big) \Rcal
&\left(\tikzDnUp\tikztauUpUp\,+\,\tikzLUpUp\tikztauUpUp\tikzCUpUp\tikztauUpUp\, + \, \tikzLUpDn\tikztauDnDn\tikzDnUp\tikztauUpUp\right)\tikzRUpUp& \\
   \label{eq:unitarity.pw.R.disc.sigma}
    &\qquad +\Rcal^\dagger\left(\tau^\dagger\dB + (\tau-\tau^\dagger)\Ladder + \tau^\dagger \dB \tau \Ladder\right).
&\tikzRUpDn\left(\tikztauDnDn\tikzDnUp\,+\,\tikztauUpUp\tikzCUpUp\tikztauUpUp\tikzLUpUp\, + \, \tikztauDnDn\tikzDnUp\tikztauUpUp\tikzLUpUp\right)&
\end{align}
\end{subequations}
The interpretation of Eq.~\eqref{eq:unitarity.pw.R} is rather clear as illustrated by the diagrams:
The two terms~\eqref{eq:unitarity.pw.R.disc.s}
represent intermediate states between $\Rcal$, either with matched isobar sub-energies or with recoupled ones.
The terms~\eqref{eq:unitarity.pw.R.disc.sigma.prime} and \eqref{eq:unitarity.pw.R.disc.sigma} are reminiscent of the unitarity relation of the ladder
with the exchange interaction attached to $\Rcal$ from the left- and the right-hand side.
The latter fact motivates the introduction of the reduced amplitude $\RcalHat(\sigma',s,\sigma)$ defined by,
\begin{align} \label{eq:our.model.ansatz}
  \tikzRUpUp && \Rcal & \equiv  (1+\Ladder\tau)\, \RcalHat\, (\tau\Ladder+1).&&
  \left(\tikzoneUpUp + \tikzLUpUp\tikztauUpUp \right)\tikzRhatUpUp\left(\tikztauUpUp\tikzLUpUp + \tikzoneUpUp\right),
\end{align}
where the structures $(1+\Ladder\tau)$ and $(\tau\Ladder+1)$ generate an infinite sum of successive attachments of exchange processes to the left and right of $\RcalHat$.
After combining Eq.~\eqref{eq:unitarity.pw.R} and Eq.~\eqref{eq:our.model.ansatz},
we find that the terms in Eq.~\eqref{eq:unitarity.pw.R.disc.sigma} and
Eq.~\eqref{eq:unitarity.pw.R.disc.sigma.prime} are eliminated due to
the unitarity property of $\Ladder$. For instance, the term~\eqref{eq:unitarity.pw.R.disc.sigma.prime} can be reduced using the following:
\begin{align*}
\big(\dB\tau + \Ladder^\dagger(\tau-\tau^\dagger) + \Ladder^\dagger \tau^\dagger \dB \tau\big) (1+\Ladder\tau) \RcalHat = (1+\Ladder\tau)\, \RcalHat -  (1+\Ladder^\dagger\tau^\dagger)\, \RcalHat.
\end{align*}
Straightforward algebraic manipulations yield
\begin{align} \label{eq:unitarity.r.1}
   \RcalHat-\RcalHat^\dagger &= \RcalHat^\dagger
   (1+\tau^\dagger \Ladder^\dagger)
   \left[
   \tau-\tau^\dagger + \tau^\dagger\dB \tau
   \right](1+\Ladder \tau )
   \, \RcalHat\\ \label{eq:unitarity.r.2}
   &= \RcalHat^\dagger
   \left[
   \tau-\tau^\dagger + \tau \Ladder \tau - \tau^\dagger\Ladder^\dagger \tau^\dagger
   \right]
   \, \RcalHat.
\end{align}
An important observation that we explore in the next section is that Eq.~\eqref{eq:unitarity.r.1} is reminiscent of two-body-unitary equations,
however, with notable differences.
The phase-space factor, proportional to $\tau - \tau^{\dagger}$, is modified by the ladder
of one-pion exchanges, \ie the terms $(1+\Ladder\tau)$ and $(1+\tau^\dagger \Ladder^\dagger)$. Moreover, it includes cross-channel terms $\tau^\dagger\dB \tau$.
It is clear from Eq.~\eqref{eq:unitarity.r.2} that a recursive solution
\footnote{
The fact that Eq.~\eqref{eq:RcalHat.Kmat} is a solution can be demonstrated as follows. First, rewrite the right-hand side of Eq.~\eqref{eq:unitarity.r.2} as
$\RcalHat^\dagger [\hat{\mathcal{O}} - \hat{\mathcal{O}}^\dagger ] \RcalHat$, with $\hat{\mathcal{O}} = \tau + \tau \Ladder \tau$.
Then, two zero additions lead to:
$$
\RcalHat^\dagger
   \left[ \hat{\mathcal{O}} - \hat{\mathcal{O}}^\dagger  \right]
    \RcalHat = \left[ \left( \RcalHat^\dagger - \Kmat^\dagger \right) \hat{\mathcal{O}} \RcalHat - \RcalHat^\dagger \hat{\mathcal{O}}^\dagger \left( \RcalHat - \Kmat \right) \right] + \left[ \Kmat^\dagger \hat{\mathcal{O}} \RcalHat - \RcalHat^\dagger \hat{\mathcal{O}}^\dagger \Kmat \right].
$$
Finally, due to the definition in Eq.~\eqref{eq:RcalHat.Kmat} and its complex
conjugation, the expression in the first square brackets vanishes.
The expression in the second brackets gives
the left-hand side of Eq.~\eqref{eq:unitarity.r.2},
when using $\Kmat^\dagger = \Kmat$ (since $\Kmat$ is real).
}
can be written down for $\RcalHat$:
\begin{equation} \label{eq:RcalHat.Kmat}
  \RcalHat = \Kmat + \Kmat(\tau + \tau\Ladder\tau)\,\RcalHat,
\end{equation}
where $\Kmat(\sigma',s,\sigma)$ is supposed to have no right-hand-cut singularities. It is a real function in the physical region, \ie $\Kmat^\dagger = \Kmat$, and plays the role of a $K$-matrix as commonly used in $2\to 2$ scattering.

Equation~\eqref{eq:RcalHat.Kmat} completes the discussion on the most broad class of models within the isobar approach. In particular, we have found that $\Tcal = \Ladder$ is a valid unitary amplitude.
Furthermore, we showed that it can be extended by a function $\Rcal$ in the form of Eq.~\eqref{eq:our.model.ansatz},
where the exchange processes dress the amplitude $\RcalHat$, while the latter obtains a simple unitarity condition, \ie Eq.~\eqref{eq:unitarity.r.2}.
Using the terminology introduced in this section, we now compare several models that are currently discussed in the literature~\cite{Mai:2017vot,Jackura:2018xnx,Hansen:2015zga}.
The present model, as well as the one from Ref.~\cite{Jackura:2018xnx}, are constructed for the partial-wave amplitude, in order to facilitate applications to resonance physics.
In contrast, Refs.~\cite{Mai:2017vot, Hansen:2015zga} keep the full scattering amplitude as a function of bachelor momenta, which is more suitable for finite volume applications.
The models of Ref.~\cite{Mai:2017vot} and Ref.~\cite{Jackura:2018xnx} explore a similar construction: the kernel function $\Ker$ includes the one-pion-exchange and the contact term.
A difference between these models can be found in the lower integration limit in Eq.~\eqref{eq:TauMultRule}.
As soon as the latter is smaller than $4\mpisq$, one obtains indeed the same value for $\Tcal - \Tcal^\dagger$ in~\eqref{eq:three.body.unitarity}, which is the only quantity constrained by unitarity.
The value $4\mpisq$ is employed in Ref.~\cite{Jackura:2018xnx}, the lower limit is set to $-\infty$ in~\cite{Mai:2017vot} (see the detailed discussion in \cite{Jackura:2018xnx}).
Treating the long-range exchanges as a separate ladder term
allowed us to focus on the short-range physics (\ie function $\RcalHat$), which is dressed with rescattering processes (see Eq.~\eqref{eq:our.model.ansatz}).
One finds that the divergence-free function $\Kcal_{\text{df,3}}$ introduced in
Ref.~\cite{Hansen:2015zga} is analogous to our $\Kmat$ function.
The divergent part of the scattering amplitude named $\mathcal{D}$ in Ref.~\cite{Hansen:2015zga} is equivalent to
the ladder function $\Ladder$ in our approach, in case only the pion-exchange process is included to $\Ker$.
Despite these similarities, the exact relation between the two approaches requires a more careful scrutiny, due to
the different symmetrization procedure and construction of the divergence-free part.

%

\section{Factorization of the rescattering}
\label{sec:factorization}

Equation~\eqref{eq:RcalHat.Kmat} gives a general solution to the three-body unitarity problem,
however, it is still a complicated integral equation.
In this section, we show how assuming factorization of final-state interactions helps to transform the constraint to an algebraic form.
Since we are interested in the direct channel dynamics only, we can expand $\RcalHat(\sigma',s,\sigma)$ in powers of $\sigma$, and $\sigma'$:
\footnote{
An alternative form of the factorization ansatz is $\RcalHat(\sigma',s,\sigma) = k_f(\sigma')\RcalFac(s)k_i(\sigma)$,
where $k_i(\sigma)$ and $k_f(\sigma)$ are arbitrary functions without the right-hand cut.
Factorization of the residues at the resonance pole would motivate such a model.
}
\begin{align} \label{eq:SmallRFactorization}
   \RcalHat(\sigma',s,\sigma) &= \RcalFac_{00}(s)+\sigma'\RcalFac_{10}(s)+\RcalFac_{01}(s)\sigma+\sigma'\RcalFac_{11}(s)\sigma+\dots,
\end{align}
where $\RcalFac_{ij}(s)$ are distinguishable coefficients.
We notice that every term factorizes in the variables $\sigma'$, $s$, and $\sigma$.
By plugging the expansion~\eqref{eq:SmallRFactorization} into Eq.~\eqref{eq:unitarity.r.1}, we obtain
an algebraic system of equations for the $\RcalHat_{ij}(s)$.
In order to demonstrate the result, we truncate the expansion at the first term,
$\RcalHat(\sigma',s,\sigma) = \RcalHat_{00}(s) \equiv \RcalHat(s)$.
We arrive at a fairly simple equation for the factorized kernel, $\RcalFac$:
\begin{align} \label{eq:unitarity.r.factorization}
   \RcalFac(s)-\RcalFac^\dagger(s) &= i\,\RcalFac^\dagger(s) \Sigma(s) \,\RcalFac(s),
\end{align}
where $\Sigma(s)$ is given by,
\begin{align} \label{eq:DressedRho.def}
  \DressedRho &\equiv
    \Kcal^\dagger(\tau-\tau^\dagger)\Kcal
    + \Kcal^\dagger \tau^\dagger\dB \tau\Kcal,
    &&
    \tikzdressedrightVertexUp\tikztauUpUp\tikzCUpUp\tikztauUpUp\tikzdressedleftVertexUp +
    \tikzdressedrightVertexUp\tikztauUpUp\tikzUpDn\tikztauDnDn\tikzdressedleftVertexDn
  \end{align}
with the correction function $\Kcal(s,\sigma)$,
\begin{align}
    \Kcal &\equiv (1+\Ladder\tau)\,1.&&
    \tikzdressedrightVertexUp = \tikzundressrightVertexUp + \tikzundressrightVertexUp\tikztauUpUp\tikzLUpUp
\end{align}
In the latter equations, we explicitly include a factor $1$ in the last term to keep the integral short-hand notations from Eq.~\eqref{eq:TauMultRule} consistent.

The structure of the unitarity equation~\eqref{eq:unitarity.r.factorization} is the same as that of the two-body unitarity in Eq.~\eqref{eq:unitarity.t}.
The two-body phase-space factor is replaced by the function $\DressedRho(s)$, which averages rescattering contributions over the three-body phase space.
As we saw before, the structure $(1+\Ladder\tau)\,1$ corresponds to an infinite number of successive attachments
of the long-range kernel on the left of source $1$. Therefore, $\Kcal (s,\sigma)$ represents the source function corrected by
the final state interaction.
We identify the two contributions in Eq.~\eqref{eq:DressedRho.def} as a direct coupling of the particles paired into the isobar term
(the first term) as well as the recoupling term (second term), where the isobar on the right is formed by including the bachelor particle from the left.
As anticipated from the general unitarity principle,
the term $\DressedRho(s)$ that determines the imaginary part of the amplitude
is equal to an integral of the resonance decay amplitude squared over the Dalitz plot.
In case the subchannel interaction is described by a single resonance,
the first term in Eq.~\eqref{eq:DressedRho.def} is an intensity of the three symmetric resonance bands, one for every subchannel,
while the second term gives the interference contribution that arises from the band overlaps.

The algebraic unitary equation is straightforward to satisfy, \eg in $K$-matrix models~\cite{Aitchison:1972ay}.
In order to keep only those singularities that are demanded by unitarity on the first sheet of the scattering amplitude, a common dispersive construction can be employed~\cite{Martin:102663}:
\begin{equation}
  \RcalFac^{-1}(s) = \KmatFac^{-1}(s) - i\tilde{\DressedRho}(s)/2,\quad \tilde{\DressedRho}(s) = \frac{s}{\pi i} \int_{9\mpisq}^{\infty} \frac{\DressedRho(s')}{s'(s'-s)} \diff s',
\end{equation}
where $\tilde{\DressedRho}(s)$ represents the dressed isobar-bachelor-loop integral (the \textit{self-energy function}) that encodes the infinite series of rescattering processes between the three particles, and
$\KmatFac(s)$ is an arbitrary function that incorporates the underlying QCD dynamics. The latter is commonly
parametrized by a sum of pole terms and polynomial components.
By construction, $\KmatFac(s)$ must be real in the physical region.
Our factorized model excludes other cross-channel processes, such as two-pion exchanges, or the presence of 3-body resonances in the $t$-channel.
However, those exchanges produce distant left-hand singularities and we suppose that the effect of them can be incorporated effectively via left-hand singularities of the function $\RcalFac(s)$, \ie included in $\Kmat$,
analogously to the customary techniques used in analyses of two-body reactions.

As one can deduce from Eq.~\eqref{eq:DressedRho.def}, the function $\Kcal$ enters in combinations such as $t(\sigma)\Kcal(s,\sigma)$,
therefore it enters solely to modify the two-body scattering amplitude.
Those corrections becomes small in the narrow-resonance limit.
For practical applications, we can identify several approximations:
\begin{itemize}
  \item An approximate-three-body-unitarity approach as employed in Ref.~\cite{Mikhasenko:2018bzm},
  which neglects the effects of final-state interactions. In this case, the model for the self-energy function is
  \begin{align} \label{eq:DressedRho.approx}
    \DressedRho_\text{approx} &=
      1\,(\tau-\tau^\dagger)\,1 + 1\,\tau^\dagger\dB \tau\,1,&
      &
      \tikzundressrightVertexUp\tikztauUpUp\tikzCUpUp\tikztauUpUp\tikzundressleftVertexUp +
      \tikzundressrightVertexUp\tikztauUpUp\tikzUpDn\tikztauDnDn\tikzundressleftVertexDn
  \end{align}
  where the first term involves an integral over the isobar mass according to Eq.~\eqref{eq:TauMultRule},
  and the second term contains two integrals.
  This model still includes a genuine three-body effect: the isobar recoupling.
  However, the corrections to the lineshape of the isobar are omitted. 
  \item The quasi-two-body approximation proposed in Ref.~\cite{Basdevant:1978tx} arises
  when also the recoupling terms are dropped. The function $\DressedRho_\text{qtb}(s)$,
  \begin{align}
    \DressedRho_\text{qtb}(s) &=
    1\,(\tau-\tau^\dagger)\,1 = \int_{4\mpisq}^{(\sqrt{s}-\mpi)^2} \rho_s(\sigma)\, \frac{t^\dagger(\sigma)\rho(\sigma)t(\sigma)}{2\pi} \,\diff \sigma
    \xrightarrow[\text{isobar}]{\text{stable}} \rho_s(m_R^2),
  \end{align}
  is approximately equal to the two-body phase space calculated for the nominal mass of the subchannel resonance, $m_R$
  above the isobar-bachelor threshold.
  $\DressedRho_\text{qtb}(s)$ continues smoothly to the region below the threshold and vanishes at the three-pion threshold.
\end{itemize}

%

\section{Connecting to Khuri-Treiman} \label{sec:SpecificModel}

We have shown that the requirement of three-body unitarity results in the necessity of the inital- and final-state interactions.
The latter was accomplished by attaching the structure $(1+\Ladder \tau)$ on both sides of the resonance amplitude.
The Khuri-Treiman approach investigates the same final-state interaction phenomena, but within the context of a production amplitude.
In this section we demonstrate that the KT framework can be brought to a form that matches the general model outlined in section~\ref{sec:Model},
with a specific ladder operator $\LadderKT$, generated from a non-trivial kernel $\KerKT$. Moreover, the KT amplitude has well-defined analytic properties due to the usage of dispersive integrals.
This makes the approach very convenient once the resonance amplitude is continued to the unphysical sheets and the properties of the resonance poles are extracted.
Factorization has been implemented customarily by giving an $s$-dependence to the subtraction constants in the KT equations~\cite{Hoferichter:2014vra}.

Experimentally, \threeToThree scattering cannot be observed.
Nevertheless, three-particle interactions appear implicitly in the context of production reactions:
in decays of heavy hadrons ($1\to 3$ process),
$2\to 3$ scattering (\eg $p\bar{p}$ and $e^+e^-$ annihilation processes),
as well as embedded in $2\to 4$ kinematics (\eg the diffractive production off
a fixed target). 
To present the KT framework, we consider a general production reaction, where the $3\pi$-state is produced from a state $\ket{\mathrm{source}}$. The $T$-matrix element for the production is parametrized as
\begin{align} \label{eq:production.amplitude}
    \tikzprodToThree\,&=\,F(s,\sigma_1,\sigma_2,\sigma_3) \\ \nonumber
                    &= \sum_{i=1}^3 \Fcal(\sigma_i,s) t(\sigma_i) = \sum_3 \tikzprodUp\tikztauUpUp
\end{align}
where we follow the same level of simplification as before, confining ourselves to a system of three identical particles with mass $m_{\pi}$.
We write the production amplitude $F$ as a sum of three partial-wave series, which are truncated to the $S$-wave in all subchannels.
Furthermore, we introduce the amputated amplitude $\Fcal(\sigma_i,s)$, similarly as in Eq.~\eqref{eq:amputated}.

The dynamics of the production amplitude is controlled by three-body unitarity.
A derivation of the constraint for $\Fcal$ proceeds in a similar way to that
for the \threeToThree amplitude derived in Appendix~\ref{app:derivation.of.unitarity}.
The KT model, in contrast, employs two-body unitarity, which is valid in the scattering domain.
By means of analytic continuation, the condition is extended to the decay region~\cite{Khuri:1960zz,Bronzan:1963mby}.
The KT constraint and the relevant three-body-unitarity conditions for $\Fcal$ read:
\begin{align}\label{eq:production.unitarity.sigma}
\text{Three-body unitarity}:&&  \Fcal(\sigma_+,s_+) - \Fcal(\sigma_-,s_-) &= \dB\tau\Fcal + \Tcal^\dagger(\tau-\tau^\dagger)\Fcal + \Tcal^\dagger\tau^\dagger\dB\tau\Fcal, \\ \label{eq:KT.discontinuity}
\text{Khuri-Treiman model}:&&  \Fcal(\sigma_+,s_+) - \Fcal(\sigma_-,s_+) &= \dB\tau\Fcal,
\end{align}
where the subscripts of the energy variables, $s$ and $\sigma$ indicate the sign of the small imaginary part added to these variables, which places the values above or below the unitarity cut in the corresponding variable.
The expectation value of the $T^\dagger$ operator sandwiched between the source
state and the three-particle state is replaced by $\Fcal^\dagger = \Fcal(s_-,\sigma_-)$,
according to the relation established in Ref.~\cite{Olive:1962xyz}.
The $s$-dependence in Eq.~\eqref{eq:KT.discontinuity} is parametric, however
a small positive imaginary part is required to continue into the decay region.
Precisely, the $s_+$ prescription was established by checking consistency of the dispersive framework with perturbation theory~\cite{Bronzan:1963mby}.
Applying the Cauchy theorem to Eq.~\eqref{eq:KT.discontinuity}, one establishes a relation between the amplitude $\Fcal(\sigma_i,s)$ and the cross-channel projections~\cite{Aitchison:1966lpz}:
\begin{align} \label{eq:Omnes.T}
  \Fcal(\sigma, s_+)=
  &\quad\Csub(\sigma,s) +
      \frac{1}{2\pi i} \int_{4\mpisq}^\infty
      \frac{\diff\sigma'}{(\sigma'-\sigma)}
      \underbrace{
        \frac{2i}{\lamHsPlus(\sigma')}\frac{1}{8\pi}
        \int_{\sigma^-(\sigma',s_{+})}^{\sigma^+(\sigma',s_{+})} \diff \sigma'' \,t(\sigma'')\Fcal(\sigma'', s)}_{
    \tikzprodDn\tikztauDnDn\tikzDnUp\,=\,\dB\tau\Fcal
      },
\end{align}
where $\Csub(\sigma,s)$ stands for the subtraction coefficients
and is often parametrized by a polynomial in $\sigma$. Its $s$-dependence is not controlled by the KT formalism.
It is important to note that since the dispersive integral in $\sigma'$ goes beyond the physical domain $4\mpisq<\sigma'<(\sqrt{s}-\mpi)^2$, the integration limits $\sigma^{\pm}(\sigma',s)$
require an analytic continuation of the expression $t(\sigma'')\Fcal(\sigma'',s)$ into the complex $\sigma''$-plane.
The solution of Eq.~\eqref{eq:Omnes.T} is fully determined by a specific source term $\Csub$,
and can be obtained via an iterative strategy starting from the zero-order perturbation $\Fcal^{(0)} = \Csub$~\cite{Albaladejo:2017hhj,Niecknig:2012sj,Guo:2015zqa,Danilkin:2014cra}.

The algebraic analysis of Eq.~\eqref{eq:Omnes.T} is convenient in the form of a so-called \textit{single-variable representation} (SVR),
which is obtained once the order of the integrals over $\sigma'$ and $\sigma''$ is swapped.
The integral over $\sigma''$ can be pulled to the front, leading to a modification of the integration limits as shown in Ref.~\cite{Aitchison:1966lpz,Pasquier:1968zz,Mikhasenko:2019phd}. The SVR becomes
\begin{align} \label{eq:main.KT.equation}
  \tikzprodUp &&
  \Fcal(\sigma,s)
    &=\Csub(\sigma,s) +
      \int_{-\infty}^{(\sqrt{s}-\mpi)^2} \frac{\diff \sigma''}{2\pi} \,\KerKT(\sigma,s,\sigma'') \tau(\sigma'') \Fcal(\sigma'',s).
  && \tikzprodCUp \,+\,\tikzprodDn\tikztauDnDn\tikzBDnUp,
\end{align}
where $\tau(\sigma'') = t(\sigma'')\rho_s(\sigma'')/3$.
We use the symbol $\KerKT(\sigma,s,\sigma'')$ to denote a function known as the Aitchison-Pasquier kernel~\cite{Aitchison:1966lpz,Pasquier:1968zz}.
This kernel and its analytic structure are rather complicated, but they have been discussed in detail in Refs.~\cite{Aitchison:1965xyz,Aitchison:1966lpz,Aitchison:1965zz,Pasquier:1968zz,Kacser:1966jmp}.
It was shown that $\KerKT$ can be written as a sum of the $S$-wave projection of the one-pion-exchange diagram, \ie $\OPE = \int \diff z/(m_\pi^2-u(s,\sigma,\sigma',z))$,
where $u(s,\sigma,\sigma',z) = \mpisq+\sigma'-(s+\mpisq-\sigma)(s+\sigma'-\mpisq)/(2s)+\lambda^{1/2}_s(\sigma)\lambda^{1/2}_s(\sigma')\,z/(2s)$,
and an extra term $\Ext$.
Using the analytic form of $\KerKT$, it can be shown that the extra terms do not contribute to the discontinuity in $s$,
such that the condition $\KerKT-\KerKT^\dagger = \dB$ holds~\cite{Pasquier:1968zz}.
Therefore, $\KerKT$ is a valid model for $\Ker$ in Eq.~\eqref{eq:ladder.definition}.
It generates a specific ladder amplitude for the $\threeToThree$ interaction,
$\LadderKT(\sigma',s,\sigma)$, which is defined by the integral equation $\LadderKT = \KerKT + \LadderKT\tau\KerKT$. We use the same short-hand notations,
understanding the lower limit of the integral in Eq.~\eqref{eq:main.KT.equation} to be $-\infty$.
The validity of the function $\LadderKT$ as a \threeToThree model was discussed in Ref.~\cite{Aitchison:1966lpz,Pasquier:1968zz}.
A known issue of the kernel $\KerKT$ has been pointed in Ref.~\cite{Pasquier:1968zz}, namely that the time-reversal symmetry of the generated ladder diagram is explicitly broken by the presence of the term $\Ext(\sigma',s,\sigma)$.
As can be shown analytically, this term is not symmetric under permutation of $\sigma$ and $\sigma'$~\cite{Kacser:1966jmp}.
However, the scale of the violation, \ie how strongly the term $\Ext$ contributes compared to the symmetric term $\OPE$, is unclear. It should be addressed in numerical studies.

The solution of the Eq.~\eqref{eq:main.KT.equation} can be expressed through $\LadderKT$ as seen by applying iterations,
\begin{align} \label{eq:production.through.ladder}
  \tikzprodUp &&
  \Fcal= \Csub+\LadderKT\tau\Csub &= (1+\LadderKT\tau)\Csub.
  &&
  \tikzprodCUp \,+\,\tikzprodCUp\tikztauUpUp\tikzLUpUp
\end{align}
The result recovers our ansatz Eq.~\eqref{eq:our.model.ansatz} for the resonance part of the model and justifies the interpretation of the structure $(1+\Ladder\tau)$ as the final-state interaction.
In practical applications of the KT framework, Eq.~\eqref{eq:Omnes.T} is used.
The total invariant mass $s$ is fixed and the function $\Csub$ is parametrized by a polynomial in $\sigma$
analogously to Eq.~\eqref{eq:SmallRFactorization}. Hence,
the solution of the Eq.~\eqref{eq:Omnes.T} is a linear combination of functions $\Kcal_i = (1+\Ladder\tau) \sigma^{i}$ with the exact same polynomial coefficients (in our simplification, $\Csub=c_0$ and therefore only $\Kcal\equiv\Kcal_0$ appears).
As we found in Sec.~\ref{sec:factorization},
the function $\Kcal$ is the only reflection of the long-range interactions that enters the resonance properties via the self-energy loop (see $\Sigma(s)$ in Eq.~\eqref{eq:DressedRho.def}).
Due to the specific long-range kernel $\KerKT$, the KT framework is not a universal approach.
Nevertheless, since it includes the dominant one-pion-exchange component,
it should cover the most important effects of the long-range interaction.
Extensive practical studies of the KT framework were performed in Ref.~\cite{Niecknig:2016fva}.
For the three-pion system with quantum numbers $I^G (J^{PC}) = 0^- (1^{--})$ and $P$-wave subchannel interaction ($\rho$ meson),
significant effects of final-state interactions were found for the energy of the system close to the $\omega$ and $\phi$ mesons.
It was demonstrated that the importance of low-energy rescattering decreases ($\Kcal \to 1$) as the total energy becomes large,
\eg $\sqrt{s}\to m_{J/\psi}$.
Using the KT-framework, we develop a good intuition and obtain a useful pictorial interpretation of the complicated three-body dynamics.

\section{Conclusions and Outlook} \label{sec:ConclusionsAndOutlook}

We have outlined the construction of general unitary models for the analysis of $3 \to 3$ processes based on the idea of a separate description of long- and short-range forces.
At first, the long-range interactions have been introduced in terms of the ladder amplitude $\Ladder$, which is a solution of the BS-type integral equation and
iterates the kernel function $\Ker$ to all orders. The function $\Ker$ has been only restricted by the condition $\Ker-\Ker^\dagger = \dB$, which is well satisfied
by \eg a one-pion-exchange process. Due to the recursive construction, the ladder amplitude, $\Ladder$ satisfies the three-particle unitarity constraint by itself. This amplitude has been extended additively by an \textit{a priori} unknown function $\Rcal$, that is used to parametrize the remaining short-range interactions.
Unitarity of the total amplitude, $\Tcal = \Ladder+\Rcal$, implies a particular requirement on the function $\Rcal$.
We have shown how the unitarity constraint can be respected by a general ansatz for $\Rcal$ introducing the resonance kernel $\RcalHat$,
dressed by the initial- and final-state interaction operators constructed from $\Ladder$.
This dressing establishes all relevant normal threshold singularities in the two-body sub-energy variables such that
the resonance kernel $\RcalHat$
is supposed to have only three-particle-threshold
singularity in the overall energy variable.
Therefore, it is possible to write a very general solution for the resonance kernel $\RcalHat$, in the form of the integral $K$-matrix parametrization.

We have imposed a factorization ansatz for the resonance kernel $\RcalHat$, that has led to a simplification of the unitarity requirement.
This results in an algebraic equation in a form which is similar to the conventional two-body-unitary condition. Therefore, all common techniques used for the two-body reaction become applicable.
Under factorization, the long-range interaction and the induced final-state rescattering are packed into the dressed isobar-bachelor loop function (the resonance self-energy function).
When the two-particle interaction
is described by a narrow resonance, the formalism
resembles the known quasi-two-body unitarity approach~\cite{Basdevant:1978tx}.

We have investigated how the known Khuri-Treiman formalism gives a specific model for the long-range kernel $\Ker$.
Due to the usage of the dispersive Khuri-Treiman approach, we obtain the advantage that the amplitude has a known and simple analytic structure. Furthermore, we have argued that the Khuri-Treiman formalism serves as a convenient tool to study the strength of rescattering corrections. As follows from our derivation, the rescattering corrections provide the only ingredient which is needed to evaluate the resonance self-energy function.

Our model is a proposal suitable for studies of the three-particle resonances.
Using the relation of the scattering amplitude and the production amplitude, the model can be applied to more complicated hadronic reactions,
as for example the process $\pi\,p\to 3\pi\,p$~\cite{Adolph:2015tqa,Akhunzyanov:2018lqa},
or in hadronic tau decays, $\tau\to3\pi\,\nu_\tau$~\cite{Asner:1999kj,Mikhasenko:2018bzm}.

The present work just illustrates the basic ideas in the rather artificial context of a restriction to $S$-waves only.
Thus, in the next step the formalism should be extended for an arbitrary value of the total angular momentum.
The system of three pions also requires a consistent isospin treatment.
An extension of this work to the coupled-channel problem is needed for further investigation of the $3\pi-K\bar{K}\pi$ coupled system
suggested to be responsible for the exotic candidate $a_1(1420)$~\cite{Adolph:2015pws,Ketzer:2015tqa,Mikhasenko:2016mox}.
We have left aside the problem of time-reversal symmetry violations in KT, which needs to be addressed in the practical cases.

\section*{Acknowledgements}
This work was supported by
the German Bundesministerium f\"{u}r Bildung und Forschung (BMBF), 
and Deutsche Forschungsgemeinschaft (DFG),
the U.S.~Department of Energy Grants No.~DE-AC05-06OR23177 
and No.~DE-FG02-87ER40365, U.S.~National Science Foundation Grant No.~PHY-1415459.
V.M. acknowledges support from the Community of Madrid through
the Programa de atracci\'on de talento investigador 2018 (Modalidad 1).


\appendix

%

\section{Derivation of the three-body-unitarity equations}\label{app:derivation.of.unitarity}
A spinless-particle state $\ket{p}$ has the customary relativistic normalization
\begin{equation} \label{eq:single.particles.normalization}
  \braket{p'}{p} = 2E(2\pi)^3\delta^3(p'-p) \equiv \tilde{\delta}_{p'p}.
\end{equation}
We define a state of two identical particles that explicitly incorporates permutation properties:
\begin{equation} \label{eq:two.particles.state}
  \ket{p_1p_2} = \frac{\ket{p_1}\ket{p_2}+\ket{p_2}\ket{p_1}}{2}.
\end{equation}
A state of three identical particles is defined by summing over all possible permutations of the particle-momenta
\begin{align} \nonumber
    \ket{p_1 p_2 p_3} &= \frac{1}{3!}\left(\ket{p_1}\ket{p_2}\ket{p_3}+\textrm{symm.}\right)\\
    \label{eq:three.particles.state}
                      &=\frac{1}{3}\sum_{a}\ket{p_{a_1}}\ket{p_{a_2}p_{a_3}} = \frac{1}{3}\sum_{a}\ket{a},
\end{align}
where we introduced a compact notation for the state $\ket{a} = \ket{p_{a_1}}\ket{p_{a_2}p_{a_3}}$,
symmetrized over the momenta $p_{a_2}$, $p_{a_3}$,
\ie $\ket{1} = \ket{p_{1}}\ket{p_{2}p_{3}}$, and the definition of the other states
$\ket{2}$ and $\ket{3}$ follow from the circular permutation.
The state $\ket{p_1 p_2 p_3}$ belongs to the direct product of three Hilbert spaces, one for every particle, $V_1\otimes V_2\otimes V_3$~\cite{Martin:1970xx}
The particles in the state cannot be distinguished by their momentum, however they belong to different subspaces, $V_Q$, $Q=1,2,3$.
The subspace index is often omitted in the equations to simplify notations.

The identity operator is defined on the basis of symmetrized states as,
\begin{align}\label{eq:identity}
  \mathbb{I} &= \int \tildediffp \ket{p_1 p_2 p_3} \bra{p_1 p_2 p_3},\quad
\tildediffp = \frac{\diff^3 p_1}{2E_1(2\pi)^3} \frac{\diff^3 p_2}{2E_2(2\pi)^3} \frac{\diff^3 p_3}{2E_3(2\pi)^3},
\end{align}
where we use a short notation for the product of integrals over the $3$-momenta.
Using the decomposition given in Eq.~\eqref{eq:three.particles.state} we can cast the expression for the identity operator in the symmetrized there-particle Hilbert space into the following convenient form
\begin{align} \nonumber
             \mathbb{I} &= \frac{1}{9}\sum_{a,b} \int \tildediffp \ket{a} \bra{b}\\
                \label{eq:identity.final.form}
             &= \frac{1}{3} \int \tildediffp \ket{1} \bra{1} + \frac{2}{3} \int \tildediffp \ket{2} \bra{3},
\end{align}
The integral is fully symmetric under permutation of particle indices. Therefore, we were able to gather terms which mix bachelor indices and those which do not. For instance,
the term $\int \tildediffp \ket{1} \bra{1}$ translates to $\int \tildediffp \ket{2} \bra{2}$ by interchanging the integration variables $p_1\leftrightarrow p_2$
and using the symmetry of the state in Eq.~\eqref{eq:two.particles.state} as follows,
\begin{equation*}
  \ket{p_1} \ket{p_2p_3}\xrightarrow[]{p_1\leftrightarrow p_2} \ket{p_2} \ket{p_1p_3} = \ket{p_2} \ket{p_3p_1}.
\end{equation*}
The Lorentz-invariant scattering matrix element $M$ is defined as the expectation value of the transition operator $T$
\begin{equation} \label{eq:three.to.three.interaction}
  \bra{p_1'p_2'p_3'} T \ket{p_1p_2p_3} = (2\pi)^4\delta^4 \left( \sum_{i=1}^3 p_i'-\sum_{i=1}^3 p_i \right)\,M,
\end{equation}
where the initial (final) particle momenta are denoted by $p_i$ ($p_i'$), $i=1,2,3$.

The unitarity condition for the $S$-matrix, \ie $S^{\dagger} S = \mathbb{I}$, transforms to a relation for the transition operator $T$: $T-T^\dagger = iT^\dagger T$. This constraint for operators in the abstract Hilbert space leads to
a constraint on the matrix elements by calculating the expectation-value with initial and final three-particle states
\begin{equation*}
  \bra{p_1'p_2'p_3'} T-T^\dagger \ket{p_1p_2p_3} = i \int \tildediffppp \bra{p_1'p_2'p_3'} T^\dagger \ket{p_1''p_2''p_3''} \bra{p_1''p_2''p_3''}T \ket{p_1p_2p_3},
\end{equation*}
where we inserted the resolution of the identity from Eq.~\eqref{eq:identity}.
The same equation holds for the states with reduced symmetry $\ket{a}$. Using Eq.~\eqref{eq:identity.final.form} and Eq.~\eqref{eq:three.particles.state}, we get
\begin{equation}  \label{eq:UnitarityReducedSymmetryStates}
  \bra{b'}T-T^\dagger\ket{a} = \frac{i}{3} \int \tildediffppp \bra{b'}T^\dagger\ket{1^{\prime \prime}}\bra{1^{\prime \prime}} T\ket{a} + \frac{2i}{3} \int \tildediffppp \bra{b'}T^\dagger\ket{2^{\prime \prime}}\bra{3^{\prime \prime}} T\ket{a}.
\end{equation}

We split the interaction operator $T$ using the connectedness principle of analytic $S$-matrix theory \cite{Eden:1966dnq}, into the (fully) \textit{connected} interaction $\Tc$ and the (partially)
\textit{disconnected} interaction $\Td$ by writing $T = \Tc + \sum_{Q=1}^3 \Td^Q$, where $\Td^Q = \mathbb{I}_Q\otimes T$ with $\mathbb{I}_Q$ being identity in the subspace $V_Q$, and $T$ is an interaction operator that acts in the remaining two-particle subspace.
The disconnected piece $\Td$ is a part of the scattering in Eq.~\eqref{eq:three.to.three.interaction} where only two particles interact while the remaining bachelor particle propagates through.
When sandwiched with the symmetrized states, the disconnected part obtains a factor of $3$ that indicates three possible ways to choose the non-interacting bachelor particle. We obtain
\begin{equation} \label{eq:expectation.disconnected}
  \bra{p_1'p_2'p_3'} \sum_{Q=1}^3 \Td^Q \ket{p_1p_2p_3} = \frac{1}{3}\sum_{a,b}\bra{b'}\Td\ket{a}
   = \frac{1}{3}\sum_{a,b}\bra{p_{b_2}'p_{b_3}'}T\ket{p_{a_2}p_{a_3}} \tilde{\delta}_{a_{1},b_{1}'}.
\end{equation}
where the bachelor state is conserved as indicated by the delta function $\tilde{\delta}_{a_{1},b_{1}'}$
(\cf Eq.~\eqref{eq:single.particles.normalization}).
Using the connectedness, Eq.~\eqref{eq:UnitarityReducedSymmetryStates} is decomposed further.
The left-hand side is separate an additive way,
since it is only linear in $T$ (as well as $T^\dagger$).
Due to the product $T^\dagger T$, the right-hand side leads to different topologies
of the types ``disconnected-disconnected'', ``disconnected-connected'' and ``connected-connected''
(see also Ref.~\cite{Fleming:1964zz,Mai:2017vot}).
A part of the ``disconnected-disconnected'' terms contains the spectator delta function and can be matched with the disconnected terms on the left-hand side. The remaining terms match the left-hand-side expression for the connected amplitude.
\begin{align} \label{eq:two.body.unitarity.app}
  \bra{b}\Td-\Td^\dagger\ket{a} &= i\int \tildediffppp \bra{b}\Td^\dagger\ket{1''}\bra{1''} \Td\ket{a}, \\
  \label{eq:three.body.unitarity.app}
  \bra{b}\Tc-\Tc^\dagger\ket{a} &= i\int \tildediffppp \big[
  \frac{1}{3}\bra{b}\Tc^\dagger\ket{1''}\bra{1''} \Tc\ket{a}  +  \frac{2}{3}\bra{b}\Tc^\dagger\ket{2''}\bra{3''}\Tc\ket{a}\\
  \nonumber
     &\hspace{2cm} + \bra{b}\Td^\dagger\ket{1''}\bra{1''} \Tc\ket{a} + 2\bra{b}\Td^\dagger\ket{2''}\bra{3''}\Tc\ket{a}\\
     \nonumber
     &\hspace{2cm} + \bra{b}\Tc^\dagger\ket{1''}\bra{1''} \Td\ket{a} + 2\bra{b}\Tc^\dagger\ket{2''}\bra{3''}\Td\ket{a}\\
     \nonumber
     &\hspace{2cm} + 6\bra{b}\Td^\dagger\ket{2''}\bra{3''}\Td\ket{a}\big].
\end{align}
We remark a similar equation can be found in Ref.~\cite{Fleming:1964zz} (see Eq.~(3.9)).

We define the partial-wave state $\ket{q_1l\lambda}$ of two particles by projecting the symmetrized state from Eq.~\eqref{eq:two.particles.state} using angular basis functions
$D_{\lambda\lambda^{\prime}}^{J}(\Omega)$~\cite{JacobWick,MartinSpearman}:
\begin{equation} \label{eq:state.lambda}
  \ket{q_1l\lambda} = \sqrt{2l+1} \int \frac{\diff \Omega_{23}}{4\pi} D_{\lambda 0}^{l}(\Omega_{23})\ket{p_2p_3} \quad\Leftrightarrow\quad
  \ket{p_2p_3} = \sum_{l\lambda}\sqrt{2l+1} \,D_{\lambda 0}^{l}(\Omega_{23}) \ket{q_1l\lambda}.
\end{equation}
\newcommand{\hfr}{\ensuremath{h_{23}}}
\newcommand{\rf}{\ensuremath{\mathrm{rf}}}
The spherical angles $\Omega_{23} = (\theta_{23}, \phi_{23})$ are defined by the direction of the momentum $\vec p_2$ in the rest frame of the pair $(23)$, \ie the frame with index $\hfr$ as shown in Fig.~\ref{fig:reference.plane}.
We stress that $z_{\hfr}$ is chosen such that spin projection $\lambda$ is conserved and becomes helicity when the state $\ket{q_1l\lambda}$ is boosted from the $\hfr$ frame to the production frame. The total momentum of this pair is denoted as $q_1$.
\begin{figure}
  \includegraphics{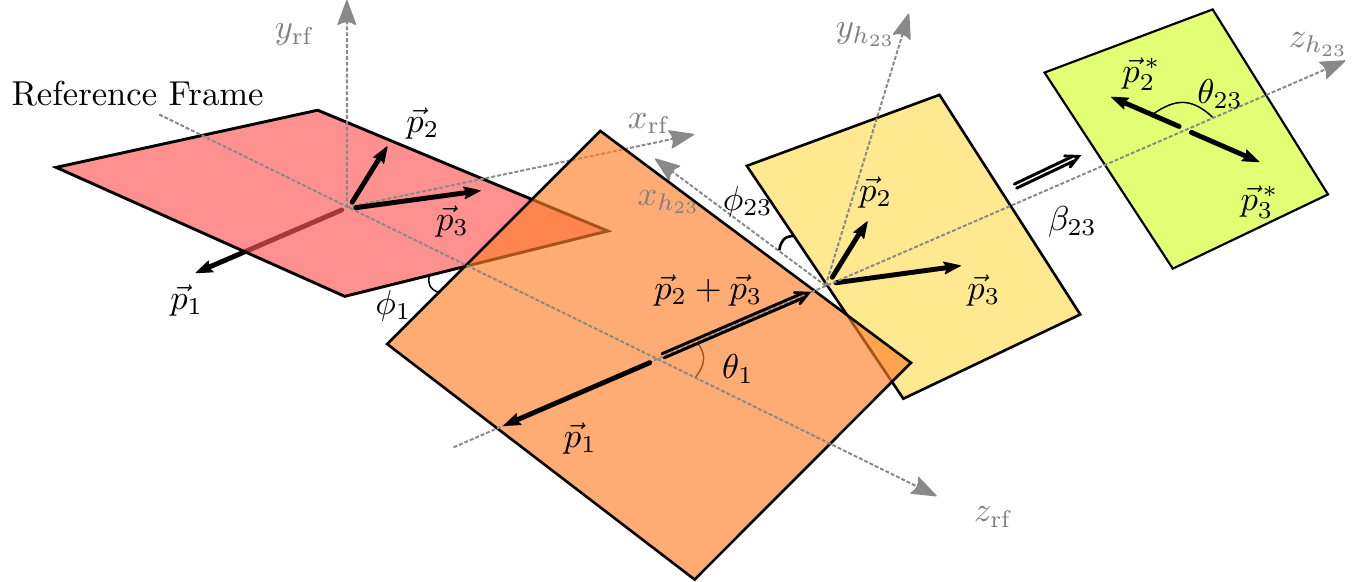}
  \caption{Definition of angles that parametrize three-body kinematics with respect to the given reference frame (see  $x_\rf$, $y_\rf$, $z_\rf$) and helicity frame of the $(23)$-particles pair (see $x_{\hfr}$, $y_{\hfr}$, $z_{\hfr}$).}
  \label{fig:reference.plane}
\end{figure}
The partial-wave-projected two-particle state is referred to as the \textit{isobar}. The
projection of the isobar-bachelor states to the total angular momentum is done analogously.
In order to define spherical angles, we consider a fixed \textit{reference} system
in the three-pion center-of-mass frame.
Often in the practical applications, the reference system can be aligned with final state particles such that
some angles vanish. However, consideration of the general case
when this system does not reply on the three-pion orientation does not complexify the equations.
\footnote{
The general case is realized when the reference frame is fixed by external kinematics, \eg Gottfried-Jackson frame~\cite{Gottfried:1964nx} for the diffractive production.
}
\begin{equation} \label{eq:state.jmlambda}
  \ket{P,jml\lambda}_1 = \sqrt{2j+1}\int\frac{\diff{\Omega_1}}{4\pi} D_{m\lambda}^{j}(\Omega_1) \,\ket{p_1}\ket{q_1l\lambda} \quad\Leftrightarrow\quad
  \ket{p_1}\ket{q_1l\lambda} = \sum_{jm}\sqrt{2j+1} \, D_{m\lambda}^{j}(\Omega_1) \,\ket{P,jml\lambda}_1,
\end{equation}
where the index $1$ of the state $\ket{P,jml\lambda}_1$ indicates the choose of the subchannel in which the partial-wave projection is performed. The total momentum of the three-particle system is $P = p_1+p_2+p_3$, $j$ and $m$ are the total orbital momentum and it's projection to the $z_\rf$ axis.
Angles $\Omega_1$ are the spherical angles of the vector $\vec q_1 = \vec p_2+\vec p_3$ in the reference system as shown in Fig.~\ref{fig:reference.plane}.

Up to this point, the derivation is general and does not invoke any model-assumptions.
For simplicity, we assume in the following that the interaction is only significant in the $S$-wave,
$j=l=0$. The two-particle $S$-wave scattering amplitude is denoted by $t$, \ie
\begin{equation*}
  \bra{q'00}\Td\ket{q00} = (2\pi)^4\delta^4(q'-q)\,t(\sigma),
\end{equation*}
where $\sigma$ is the two-particle invariant mass squared: $\sigma = q^2$.
The expectation value of the disconnected operator reads from Eq.~\eqref{eq:expectation.disconnected} and Eq.~\eqref{eq:state.lambda},
\begin{equation} \label{eq:Td.partial.waves}
  \bra{1'}\Td\ket{1} = (2\pi)^4\delta^4(q_1'-q_1) \tilde\delta_{p_1'p_1} \, t(\sigma_1).
\end{equation}
Substituting Eq.~\eqref{eq:Td.partial.waves} into Eq.~\eqref{eq:two.body.unitarity.app}, we obtain the standard two-body-unitarity relation
\begin{equation} \label{eq:two.body.unitarity.app.s.wave}
  t(\sigma)-t^\dagger(\sigma) = i t^\dagger(\sigma) \rho(\sigma) t(\sigma) \theta(\sigma-4\mpisq).
\end{equation}
The phase space, $\rho(\sigma)$ is obtained by simplifying integrals on the right-hand side of Eq.~\eqref{eq:two.body.unitarity.app} as follows
\begin{align*}
  &\rho(\sigma)\theta(\sigma-4\mpisq) = \int \tildediffppp (2\pi)^4\delta^4(q_1''-q_1) \tilde\delta_{p_1''p_1} \\
  &\qquad\quad=
  \int \frac{\diff^3 p_1''}{2E_1''(2\pi)^3} \frac{\diff^3 p_2''}{2E_2''(2\pi)^3} \frac{\diff^3 p_3''}{2E_3''(2\pi)^3} (2\pi)^4\delta^4(q_1''-q_1)\,2E_1 (2\pi)^3 \delta^3(p_1''-p_1) \\
  &\qquad\quad= \int \frac{\diff^3 p_2''}{2E_2''(2\pi)^3} \frac{\diff^3 p_3''}{2E_3''(2\pi)^3} (2\pi)^4\delta^4(q_1''-q_1) = \frac{1}{8\pi} \frac{2|\vec p_2^{\,*}|}{\sqrt{\sigma}}\theta(\sigma-4\mpisq)\\
  &\qquad\quad= \frac{1}{8\pi}\sqrt{1-\frac{4\mpisq}{\sigma}}\,\theta(\sigma-4\mpisq).
\end{align*}
The resulting expression is the two-body phase space~\cite{PeskinSchroeder,BycklingKajantie:1971},
calculated at the center-of-mass frame of the pair $(23)$ with $|\vec p_2^{\,*}|$ being the break-up momentum $|\vec p_2^{\,*}| = \sqrt{\sigma/(4\mpisq)-1}$.
The general unitarity equation, Eq.~\eqref{eq:three.body.unitarity.app} is transformed to a condition for the partial-wave-projected amplitude, once we replace $\ket{a}$ and $\bra{b}$ in the initial and the final states by the projected states $\ket{P,0000}_a$ and $_b\bra{P',0000}$, introduced in Eq.~\eqref{eq:state.jmlambda}. When replacing
we can drop the subchannel index of the in- and out- states $\ket{P,0000}$ consistently at both sides
due to the freedom in the choose of $a$ and $b$.
The partial-wave-projected connected isobar amplitude is defined by
\begin{align} \label{eq:Tc.partial.waves}
  \bra{P',0000}T_c\ket{P,0000} &= (2\pi)^4\delta^4(P'-P)\,T(\sigma',s,\sigma),
\end{align}
and depends at most on three invariant variables:\footnote{
By counting degrees of freedom one finds that the \threeToThree scattering amplitude depends on $8$ kinematic variables.
The partial-wave projections to the two-particle subchannels, given in Eq.~\eqref{eq:state.lambda} for the initial and the final states, replace four continuous angular variables by the discrete indices $l\lambda$ and $l'\lambda'$.
The partial-wave projection to the total angular momentum states defined in Eq.~\eqref{eq:state.jmlambda}
becomes equivalent to the single integration over the scattering angle between the isobar directions on the initial and final state due to the choice of the reference-frame that assigns the discrete index $j$ to the amplitude. The remaining partial-wave amplitude depends on three variables as introduced in Eq.~\eqref{eq:Tc.partial.waves}.
}
$s$ is the total invariant mass square of the three-particle system; the variables $\sigma$ and $\sigma'$ denote the squared masses of the isobars in the initial and final states, respectively.

In order to proceed with the unitarity equation for the fully projected states $\ket{P,0000}$,
the structures on the right side of Eq.~\eqref{eq:three.body.unitarity.app} have to be expressed through already introduced amplitudes $T(\sigma',s,\sigma)$ and $t(\sigma)$.
It reads,
\begin{align}
  \bra{1''}\Tc\ket{P,0000} &= (2\pi)^4\delta^4(P''-P)\,T(\sigma_1'',s,\sigma).\\
  \bra{1''}\Td\ket{P,0000} &=
  (2\pi)^4\delta^4(P''-P)\,(2\pi)\delta(\sigma_1''-\sigma)\,\frac{t(\sigma)}{\rho_s(\sigma)},
\end{align}
where there is no angular dependence on the right-side of equations since
only $S$-wave interaction is considered.
To obtain the last equation, we related the states
on the both sides to the $\ket{ql\lambda}$ using Eq.~\eqref{eq:state.lambda} and Eq.~\eqref{eq:state.jmlambda} and exploring a property of $\delta$-function,
\begin{equation*}
  2E_{q_1}(2\pi)^3 \delta^3(p_1'' - p_1) = (4\pi)\delta(\Omega_1''-\Omega_1)\,(2\pi)\delta(\sigma_1''-\sigma_1)\,\frac{1}{\rho_s(\sigma_1)}.
\end{equation*}

The integrals $\int \tildediffppp$ in Eq.~\eqref{eq:three.body.unitarity.app} are combined with the $\delta^4$-functions in Eq.~\eqref{eq:Tc.partial.waves} and yield the standard three-body phase space, $\diff \Phi_3$.
\begin{align} \label{eq:three.body.phase.space}
    \diff \Phi_3 &= \frac{\diff^3 p_1''}{2E_1''(2\pi)^3} \frac{\diff^3 p_2''}{2E_2''(2\pi)^3} \frac{\diff^3 p_3''}{2E_3''(2\pi)^3} (2\pi)^4\delta^4(P''-P) \\
    \label{eq:three.body.phase.space.form1}
    & = \frac{\diff \sigma_1''}{2\pi}\,\rho(\sigma_1'') \rho_s(\sigma_1'') \frac{\diff \Omega_1}{4\pi}\frac{\diff \Omega_{23}}{4\pi}\,
    \theta^+(\phi(\sigma_2,s,\sigma_3))\\
    \label{eq:three.body.phase.space.form2}
    & = \frac{\diff \sigma_2'' \diff \sigma_3''}{2\pi(8\pi)^2 s} \frac{\diff \Omega_2}{4\pi}\frac{\diff \phi_{31}}{2\pi}\,
    \theta^+(\phi(\sigma_2,s,\sigma_3)),
\end{align}
where $\lamHs(\sigma(\sigma_2,s,\sigma_3)) = \lambda^{1/2}(s,\sigma,\mpisq)$, $\rho_s(\sigma) = \lamHs(\sigma)/(8\pi s)$ and $\lambda(x,y,z) = x^2+y^2+z^2-2xy-2yz-2zx$ is the \Kallen function.
We introduced a function $\theta^+(\phi(\sigma_2,s,\sigma_3))$ that restricts the variable ranges to physical domain of the three-body phase space.
\begin{equation}\label{eq:Heaviside.plus}
  \theta^+(\phi(\sigma_2,s,\sigma_3)) \equiv \theta(\phi(\sigma_2,s,\sigma_3)) \,\theta(s-9\mpisq) \theta(\sigma_2-4\mpisq)\theta(\sigma_3-4\mpisq)
\end{equation}
with $\phi$ being the Kibble function, $\phi(\sigma_2,s,\sigma_3) = \sigma_2\sigma_3(3\mpisq+s-\sigma_2-\sigma_3)-\mpisq(s-\mpisq)^2$.
The Heaviside functions $\theta(\sigma_i-4\mpisq)$, $i=2,3$ cut off the scattering domains on the Mandelstam plane for which
$\phi$ is still positive.
For the terms that arise from the insertion of $\ket{1''}\bra{1''}$
on the right-hand side of Eq.~\eqref{eq:three.body.unitarity.app}, both amplitudes $\Tcal^\dagger$ and $\Tcal$ depend on the same variable $\sigma_1$, therefore four of five integrals in the phase space can be solved
using the representation in Eq.~\eqref{eq:three.body.phase.space.form1}.
For the terms from the $\ket{2''}\bra{3''}$-intermediate state, the form of the phase space given in Eq.~\eqref{eq:three.body.phase.space.form2} is more appropriate. In this case, three angular integrals can be solved analytically.

The resulting three-body-unitarity equation in the partial-wave-projected form reads:
\begin{subequations} \label{eq:unitarity.pw}
\begin{align} \label{eq:pw.three.body.unitarity.line1}
  T(\sigma',s,\sigma)-T^\dagger(\sigma',s,\sigma) &=
  \frac{i}{3}\,\int_{4\mpisq}^{(\sqrt{s}-\mpi)^2} \frac{\diff \sigma''}{2\pi}\, T^\dagger(\sigma',s,\sigma'') \rho(\sigma'')\rho_s(\sigma'') T(\sigma'',s,\sigma)\\
  &\hspace{5mm}+  \frac{2i}{3}\,\frac{1}{(8\pi)^2}\iint_{\phi(\sigma_2'',s,\sigma_3'')>0} \frac{\diff \sigma_2''\diff \sigma_3''}{2\pi s}\, T^\dagger(\sigma',s,\sigma_2'') T(\sigma_3'',s,\sigma)\\
  \label{eq:pw.three.body.unitarity.line2}
     &\hspace{5mm} + i\, t^\dagger(\sigma')\rho(\sigma') T(\sigma',s,\sigma) +
     2i \frac{t^\dagger(\sigma')}{\lamHs(\sigma')} \frac{1}{8\pi}\int_{\sigma^-(\sigma',s)}^{\sigma^+(\sigma',s)} \diff\sigma_3' T(\sigma_3',s,\sigma) \\
     \label{eq:pw.three.body.unitarity.line3}
     &\hspace{5mm} + i\, T^\dagger(\sigma',s,\sigma) \rho(\sigma) t(\sigma) +
     2i\, \frac{t(\sigma)}{\lamHs(\sigma)} \frac{1}{8\pi}\int_{\sigma^-(\sigma,s)}^{\sigma^+(\sigma,s)} \diff\sigma_2 T^\dagger(\sigma',s,\sigma_2) \\
     \label{eq:pw.three.body.unitarity.line4}
     &\hspace{5mm} + 6i\,\frac{2\pi s\, t^\dagger(\sigma')t(\sigma)}{\lamHs(\sigma')\lamHs(\sigma)} \,\theta^+(\phi(\sigma',s,\sigma)),
\end{align}
\end{subequations}

It is convenient to define an amputated amplitude $\Tcal(\sigma',s,\sigma)$ for which we remove the last two-body interaction $t(\sigma)$ from both sides.
\begin{equation*}
  T(\sigma',s,\sigma) = t(\sigma')\Tcal(\sigma',s,\sigma)t(\sigma).
\end{equation*}
The function $\Tcal(\sigma',s,\sigma)$ still has dependencies on all variables, however the unitarity equation for $\Tcal$ is simpler.
The left-hand side of Eq.~\eqref{eq:unitarity.pw} admits the following terms grouping,
\begin{align} \nonumber
  T(\sigma',s,\sigma) - T^\dagger(\sigma',s,\sigma) &=
  	 \left[t(\sigma')-t^\dagger(\sigma')\right]\Tcal(\sigma',s,\sigma)t(\sigma) + \\
  	 \nonumber
  	&\qquad +t^\dagger(\sigma')\left[\Tcal(\sigma',s,\sigma)-\Tcal^\dagger(\sigma',s,\sigma)\right]t(\sigma)\\
  	&\qquad +t^\dagger(\sigma')\Tcal^\dagger(\sigma',s,\sigma)\left[t(\sigma)-t^\dagger(\sigma)\right].
  	\label{eq:amputated.app}
\end{align}
The differences in the first and third line can be exactly matched to the first terms in Eq.~\eqref{eq:pw.three.body.unitarity.line2} and Eq.~\eqref{eq:pw.three.body.unitarity.line3} due to the unitarity of the amplitude $t$ in Eq.~\eqref{eq:two.body.unitarity.app.s.wave}.
The unitarity equation for the amputated amplitude reads:
\begin{align}\label{eq:unitary.pw.amputated}
  \Tcal(\sigma',s,\sigma)-\Tcal^\dagger(\sigma',s,\sigma) &=
  \frac{i}{3}\,\int_{4\mpisq}^{(\sqrt{s}-\mpi)^2}  \frac{\diff \sigma''}{2\pi} \Tcal^\dagger(\sigma',s,\sigma'') \,t(\sigma'')t^\dagger(\sigma'') \,
  \rho(\sigma'')\rho_s(\sigma'')\,\Tcal(\sigma'',s,\sigma)  \\ \nonumber
     &\quad + \frac{2i}{3}\,\frac{1}{(8\pi)^2}\iint_{\phi(\sigma_2'',s,\sigma_3'')>0}  \frac{\diff \sigma_2''\diff \sigma_3''}{2\pi s}\,
        \Tcal^\dagger(\sigma',s,\sigma_{2}'')\,t^\dagger(\sigma''_{2})t(\sigma''_{3}) \Tcal(\sigma_{3}'',s,\sigma)\\ \nonumber
     &\quad + 2i\, \frac{1}{\lamHs(\sigma')} \frac{1}{8\pi}\int_{\sigma^-(\sigma',s)}^{\sigma^+(\sigma',s)} \diff\sigma_3'\,t(\sigma_{3}') \Tcal(\sigma_{3}',s,\sigma) \\ \nonumber
     &\quad + 2i\, \frac{1}{\lamHs(\sigma)} \frac{1}{8\pi}\int_{\sigma^-(\sigma,s)}^{\sigma^+(\sigma,s)} \diff\sigma_2\,t^\dagger(\sigma_{2}) \Tcal^\dagger(\sigma',s,\sigma_{2})\\ \nonumber
     &\quad + 6i\,\frac{2\pi s}{\lamHs(\sigma')\lamHs(\sigma)} \,\theta^+(\phi(\sigma',s,\sigma)),
\end{align}
where $\sigma^\pm(s,\sigma) = (s+3\mpisq-\sigma)/2 \pm \lamHs(\sigma)\lambda^{1/2}(\sigma)/(2\sigma)$.
The last term contains the Heaviside function for which a non-zero domain in $\sigma$, $\sigma'$ variables corresponds to a Dalitz plot region
for a decay of system of invariant mass $\sqrt{s}$, $\sqrt{s}>3\mpi$, to three particles of mass $\mpi$.
Since $\sigma^\pm(s,\sigma)$ are exactly the borders of this physical region, we insert the Heaviside function
$\theta^+(\phi(\sigma',s,\sigma))$ under the integrals and
unify all integration ranges to $[4\mpisq,(\sqrt{s}-\mpi)^2]$.
\begin{align} \label{eq:unitary.with.B.terms}
  \Tcal(\sigma',s,\sigma)-\Tcal^\dagger(\sigma',s,\sigma) &=
     \int_{4\mpisq}^{(\sqrt{s}-\mpi)^2} \frac{\diff \sigma''}{2\pi}
     \Tcal^\dagger(\sigma',s,\sigma'') \,(\tau(\sigma'')-\tau^\dagger(\sigma''))\,\Tcal(\sigma'',s,\sigma)  \\ \nonumber
     &\quad +
     \iint_{4\mpisq}^{(\sqrt{s}-\mpi)^2} \frac{\diff \sigma_2''}{2\pi} \frac{\diff \sigma_3''}{2\pi}\,
        \Tcal^\dagger(\sigma',s,\sigma_{2}'') \tau^\dagger(\sigma''_{2})
        \dB(\sigma_{2}'',s,\sigma_{3}'') \tau(\sigma''_{3}) \Tcal(\sigma_{3}'',s,\sigma)\\ \nonumber
     &\quad + \int_{4\mpisq}^{(\sqrt{s}-\mpi)^2} \frac{\diff\sigma_3'}{2\pi}\,\dB(\sigma',s,\sigma_{3}) \tau(\sigma_{3}') \Tcal(\sigma_{3}',s,\sigma) \\ \nonumber
     &\quad + \int_{4\mpisq}^{(\sqrt{s}-\mpi)^2} \frac{\diff\sigma_2}{2\pi}\, \Tcal^\dagger(\sigma',s,\sigma_{2}) \tau^\dagger(\sigma_{2}) \dB(\sigma_{2},s,\sigma)\\ \nonumber
     &\quad + \dB(\sigma',s,\sigma),
\end{align}
where $\tau(\sigma) = t(\sigma)\rho_s(\sigma)/3$, and $\dB(\sigma',s,\sigma) = 2\pi i\times 6s/(\lambda_s^{1/2}(\sigma)\lambda_s^{1/2}(\sigma'))\theta^+(\phi(\sigma',s,\sigma))$.

\bibliographystyle{apsrev4-1}
\bibliography{threebody}
\end{document}

%% file: review_tool.tex
\usepackage{mfirstuc} 
\usepackage{todonotes} 
\newcommand{\addReviewer}[2]{
  \expandafter\newcommand\csname #1\endcsname[1]{{\textbf{ \color{#2} \capitalisewords{#1}:\,##1}}}
  \expandafter\newcommand\csname #1cor\endcsname[2]{{\color{#2} \capitalisewords{#1}:\,\st{##1}{\textbf{##2}}}}
  \expandafter\newcommand\csname #1color\endcsname{#2}
  \expandafter\newcommand\csname #1todo\endcsname[1]{{\todo[inline,color=white!70!#2, caption={}]{\textbf{\capitalisewords{#1}}: ##1}}}
}

\usepackage{tikz, marginnote} 
\newcommand{\checkedby}[1]{
\ifdefined\CROSSCHECKS
  \marginnote{
    \begin{tikzpicture}
      \foreach \x [count=\xi] in {#1} {
         \node[shape=circle,inner sep=0mm,
         minimum size=2mm,
         fill=\csname \x color\endcsname] at (\xi*3mm,0) {};
       }
    \end{tikzpicture}
  }
\else
\fi
}

\usepackage{soul,color,color}
\definecolor{chromeyellow}{rgb}{1.0, 0.65, 0.0}
\definecolor{DodgeBlue}{rgb}{0.118, 0.565,1.000}
\definecolor{asparagus}{rgb}{0.53, 0.66, 0.42}
\definecolor{cadmiumgreen}{rgb}{0.0, 0.42, 0.24}
\definecolor{blue(ryb)}{rgb}{0.01, 0.28, 1.0}
\definecolor{turquoiseblue}{rgb}{0.02, 0.55, 0.55}

\addReviewer{bernhard}{orange}
\addReviewer{adam}{red}
\addReviewer{ale}{chromeyellow}
\addReviewer{vincent}{brown}
\addReviewer{misha}{cadmiumgreen}
\addReviewer{jannes}{blue}
\addReviewer{cesar}{magenta}
\addReviewer{tomasz}{cyan}
\addReviewer{miguel}{DodgeBlue}
\addReviewer{andrew}{purple}
\addReviewer{arkaitz}{asparagus}
\addReviewer{yannick}{blue(ryb)}
\addReviewer{mathias}{turquoiseblue}

%% file: notations.tex
\newcommand\mpi{m_\pi}
\newcommand\mpisq{m_\pi^2}

\newcommand{\threeToThree}{\ensuremath{3\!\rightarrow\!3}\xspace}

\newcommand{\Omnes}{Omn\`{e}s'\xspace}

\newcommand{\ket}[1]{\left|#1\right\rangle}
\newcommand{\bra}[1]{\left\langle#1\right|}
\newcommand{\braket}[2]{\big\langle#1\big|#2\big\rangle}

\newcommand{\diff}{\mathrm{d}}

\newcommand{\Tc}{T_\mathrm{c}}
\newcommand{\Td}{T_\mathrm{d}}
\newcommand{\tildediffp}{{(\tilde\diff p)}}
\newcommand{\tildediffppp}{{(\tilde\diff p'')}}

\newcommand{\ie}{\textit{i.e.}\ }

\newcommand{\eg}{\textit{e.g.}\ }
\newcommand{\cf}{\textit{cf.}\ }

\newcommand{\Kallen}{K\"{a}ll\'{e}n\xspace}

\newcommand{\Csub}{\widehat{\mathcal{C}}}
\newcommand{\Tcal}{\mathcal{T}}
\newcommand{\Bcal}{\mathcal{B}}
\newcommand{\Lcal}{\mathcal{L}}
\newcommand{\Kcal}{\mathcal{K}}
\newcommand{\Rcal}{\mathcal{R}}
\newcommand{\Fcal}{\mathcal{F}}
\newcommand{\Ecal}{\mathcal{E}}
\newcommand{\RcalHat}{\widehat{\Rcal}}

\newcommand{\Kmat}{\mathcal{X}}

\newcommand{\RcalFac}{\widehat{\mathcal{R}}}

\newcommand{\KmatFac}{\mathcal{X}}

\newcommand{\Ladder}{{\Lcal}} 
\newcommand{\OPE}{\Ecal}
\newcommand{\Ext}{\Ecal_{23}}
\newcommand{\Ker}{\Bcal}
\newcommand{\DressedRho}{\Sigma} 

\newcommand{\KerKT}{\Bcal_0}
\newcommand{\LadderKT}{{\Ladder_0}}

\newcommand{\dB}{\mathcal{D}}

\newcommand{\lamHs}{\lambda_s^{1/2}}
\newcommand{\lamHsPlus}{\lambda_{s_{+}}^{1/2}}

%% file: mytikzset.tex
\usetikzlibrary{positioning,arrows,shapes,patterns}
\usetikzlibrary{decorations.pathmorphing}
\usetikzlibrary{decorations.markings}
\usetikzlibrary{decorations.pathreplacing}
\usetikzlibrary{calc}
\usetikzlibrary{patterns}
\usetikzlibrary{decorations.markings}
\usetikzlibrary{petri}

\tikzset{
  vector/.style={thick,double,draw=black, postaction={decorate},
    decoration={markings,mark=at position .6 with {\arrow[black,scale=0.4]{triangle 45}}}},
  axial/.style={thick,double,densely dashed,draw=black, postaction={decorate},
    decoration={markings,mark=at position .6 with {\arrow[black,scale=0.4]{triangle 45}}}},
  gluon/.style={decorate, draw=black,
    decoration={coil,aspect=0.3,segment length=5pt,amplitude=3pt}},
  pseudo/.style={thick, dashed, draw=black, postaction={decorate},
    decoration={markings,mark=at position .6 with {\arrow[red,scale=0.5]{triangle 45}}}},
  scalar/.style={thick,draw=black, postaction={decorate},
    decoration={markings,mark=at position .6 with {\arrow[black,scale=0.5]{triangle 45}}}},
  cut/.style={draw=black, decorate, decoration={snake, segment length=1mm, amplitude=0.2mm}}
}

%% file: tikzdiagrams.tex
\newcommand{\var}{
\def\xt{3mm}   
\def\xm{1.5mm}
\def\yt{1.5mm}
\def\dyt{0.3mm}
\def\scl{1}
\def\Rad{2.4mm}
\def\rad{0.8mm}
}

\global\newcommand{\addCommand}[1]{
\expandafter\newcommand\csname tikz#1\endcsname{
  \var
  \begin{tikzpicture}[baseline=-0.8mm,scale=\scl,
      Double/.style={
          to path={
            ($(\tikztostart)!\dyt!90: (\tikztotarget)$) -- ($(\tikztotarget)!\dyt!270:(\tikztostart)$)
            ($(\tikztostart)!\dyt!270:(\tikztotarget)$) -- ($(\tikztotarget)!\dyt!90: (\tikztostart)$)
          }
        },
        Triple/.style={
            to path={
              ($(\tikztostart)!1.3*\dyt!90: (\tikztotarget)$) -- ($(\tikztotarget)!1.3*\dyt!270:(\tikztostart)$)
              (\tikztostart) -- (\tikztotarget)
              ($(\tikztostart)!1.3*\dyt!270:(\tikztotarget)$) -- ($(\tikztotarget)!1.3*\dyt!90: (\tikztostart)$)
            }
          }]
    \expandafter\csname#1\endcsname
\end{tikzpicture}\hspace{-0.3mm}
}}

\addCommand{prodUp}
\addCommand{prodCUp}
\addCommand{prodDn}
\addCommand{prodToThree}
\addCommand{diagThreeToThree}

\addCommand{conUpUp}
\addCommand{conUpUpFULL}
\addCommand{conUpDn}
\addCommand{conUpDnFULL}
\addCommand{conDnUp}
\addCommand{conDnUpFULL}
\addCommand{tauUpUp}
\addCommand{oneUpUp}
\addCommand{tauDnDn}
\addCommand{DnUp}
\addCommand{UpDn}
\addCommand{BDnUp}
\addCommand{BUpDn}
\addCommand{CUpUp}
\addCommand{RUpUp}
\addCommand{LUpUp}
\addCommand{RUpDn}
\addCommand{LUpDn}
\addCommand{RhatUpUp}
\addCommand{dressedrightVertexUp}
\addCommand{dressedleftVertexUp}
\addCommand{dressedleftVertexDn}
\addCommand{undressrightVertexUp}
\addCommand{undressleftVertexUp}
\addCommand{undressleftVertexDn}
